# Influence of Impurity Scattering on the Conductance Anomalies of Quantum Point Contacts with Lateral Spin-Orbit Coupling

J. Wan, M. Cahay, *Fellow, IEEE*, P. P. Das, and R. S. Newrock

*Abstract*—We have recently shown that asymmetric lateral spin orbit coupling (LSOC) resulting from the lateral in-plane electric field of the confining potential of a side-gated quantum point contact (QPC) can be used to create a strongly spin-polarized current by purely electrical means[1] *in the absence of applied magnetic field.* Using the non-equilibrium Green function formalism (NEGF) analysis of a small model QPC[2], three ingredients were found to be essential to generate the strong spin polarization: an asymmetric lateral confinement, a LSOC induced by the lateral confining potential of the QPC, and a strong electron-electron (e-e) interaction.

In this paper, NEGF is used to study how the spin polarization is affected by the presence of impurities in the central portion of the QPC. It is found that the number, location, and shape of the conductance anomalies, occurring below the first quantized conductance plateau ($G_0=2e^2/h$), are strongly dependent on the nature (attractive or repulsive) and the locations of the impurities. We show that the maximum of the conductance spin polarization is affected by the presence of impurities. For QPCs with impurities off-center, a conductance anomaly appears below the first integer step even for the case of symmetric bias on the two side gates. These results are of practical importance if QPCs in series are to be used to fabricate all-electrical spin valves with large ON/OFF conductance ratio.

## I. INTRODUCTION

FOR more than a decade, there have been many experimental reports of anomalies appearing at non-integer values of the quantized conductance $G_0$ in the ballistic conductance of QPCs based on GaAs. These include the observation of an anomalous plateau at $G \simeq 0.5G_0$[3-6] and the well-known "0.7 structure"[7]. The majority of the theoretical models link them to spontaneous spin polarization in the QPC[8-10]. Recently, we have used a NEGF approach to study in detail the ballistic conductance of asymmetrically biased side-gated quantum point contacts (QPCs) in the presence of lateral spin-orbit coupling and electron-electron interactions. We performed simulations for a wide range of QPC dimensions and gate bias voltage[11].

Manuscript received June 15, 2011. This work was supported by NSF Awards ECCS 0725404 and 1028483.
J. W. is with the School of Electronics and Computing Systems, University of Cincinnati, Cincinnati, OH 45221 USA (e-mail: wanjunjun@yahoo.com).
M. C. is with the School of Electronics and Computing Systems, University of Cincinnati, Cincinnati, OH 45221 (phone: 513-556-4754; fax: 513-556-7326; e-mail: marc.cahay@ uc.edu).
P. P. D is with the School of Electronics and Computing Systems, University of Cincinnati, Cincinnati, OH 45221 USA (e-mail: ppdas17@gmail.com).
R. S. N. is with the Physics Department, University of Cincinnati, Cincinnati, OH 45221 (e-mail: newrockr@ucmail.uc.edu).

Various conductance anomalies were predicted below the first quantized conductance plateau ($G_0=2e^2/h$); these occur due to spontaneous spin polarization in the narrowest portion of the QPC. We have found that the number of conductance anomalies increases with the aspect ratio (length/width) of the QPC constriction. These anomalies are fingerprints of spin textures in the narrow portion of the QPC[11].

In this paper, we investigate how these results are affected by the presence of impurities in the narrow portion of the QPC. In fact, since the early 1990s, there has been a considerable amount of work studying the influence of impurity scattering on the quantized conductance plateaus of QPCs[12-17].

We investigate here the influence of impurities on the conductance of GaAs QPCs created by side-gates in the presence of LSOC. We also study the influence of impurities on the amount of spin polarization in the QPC and its effect on the conductance anomalies. We consider the effects of attractive and repulsive scatterers located in the narrow portion of the QPC. The latter is modelled as depicted in Fig.1.

One of the main ingredients to generate spin polarization in the central portion of a QPC is the creation of an asymmetric potential profile in the channel. It is therefore expected that off-center impurities can lead to such an asymmetry even for the case of identical bias voltage on the two side gates. In this case, our simulations predict that conductance anomalies can be observed in an otherwise perfectly symmetric QPC due to unwanted impurities in the channel.

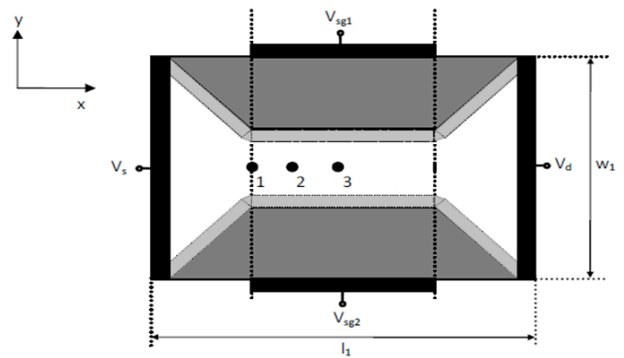

Fig.1: Schematic of the QPC configuration used in the numerical simulations. The width and length of the narrow portion of the QPC is equal to $w_2$ and $l_2$, respectively. In the simulations, we used $w_2$, $l_2$, $w_1$, $l_1$ = 16, 32, 48, and 64 nm, respectively. The impurity locations 1,2,3,4,5 correspond to the coordinates $y_1=w_1/2$ and $x_1$ = 16, 24, 32, 40, and 48 nm, respectively.

## II. NUMERICAL SIMULATIONS

The conductance through the QPC was calculated using a NEGF method under the assumption of ballistic transport[4,18] and the Green's function is given by

$$G(E) = (EI - H - \Sigma_S - \Sigma_D - \Sigma_{int})^{-1}, \quad (1)$$

where $\Sigma_S$ and $\Sigma_D$ are the self-energy terms representing the coupling of the source and drain contacts[4] and $\Sigma_{int}$ is the electron-electron interaction self-energy. For spin $\sigma$ (where $\sigma = \uparrow, \downarrow$), it is given by

$$\Sigma_{int}^\sigma(r,r') = U_H(r)\delta(r-r') + \Sigma_F^\sigma(r,r'), \quad (2)$$

where $U_H(r)$ is the Hartree potential which has the same value for both spins and includes a sum over both spin direction

$$U_H(r) = \sum_\sigma \iint \frac{1}{2\pi} G_\sigma^n(r,r',E) V_{int}(r,r') dr' dE, \quad (3)$$

where $V_{int}(r,r')$ is the e-e interaction and $G^n(r,r',E)$ is the electron correlation function. In Eq. (2), the exchange potential energy is calculated as follows

$$\Sigma_F^\sigma(r,r') = -\int \frac{1}{2\pi} G_\sigma^n(r,r',E) V_{int}(r,r') dE. \quad (4)$$

The exchange potential normally is non-local and depends on the spin orientation.

At the interface between the rectangular region of size $w_2 \times l_2$ and vacuum, the conduction band discontinuities at the bottom and the top interface were modelled, respectively, as

$$\Delta E_c(y) = \frac{\Delta E_c}{2}\left[1 + \cos\frac{\pi}{d}\left(y - \frac{w_1 - w_2}{2}\right)\right], \quad (5)$$

and

$$\Delta E_c(y) = \frac{\Delta E_c}{2}\left[1 + \cos\frac{\pi}{d}\left(\frac{w_1 + w_2}{2} - y\right)\right], \quad (6)$$

to achieve a smooth conductance band change, where $d$ was selected to be in the nm range to represent a gradual variation of the conduction band profile from the inside of the quantum wire to the vacuum region. A similar grading was also used along the walls going from the wider part of the channel to the central constriction of the QPC (Fig.1). This gradual change in $\Delta E_c(y)$ is responsible for the LSOC that triggers the spin polarization of the QPC in the presence of an asymmetry in $V_{sg1}$ and $V_{sg2}$. The parameter $d$ appearing in Eqns. (5) and (6) was set equal to 1.6 nm.

In our simulations, we model the e-e interaction $V_{int}(r,r')$ using the following *non-local* 2D screened potential[19]:

$$V_{int}(r,r') = \frac{e^2}{4\pi\varepsilon_0\varepsilon_r}\left\{\frac{1}{|r-r'|} - \frac{\pi}{2\lambda}\left(H_0\left(\frac{|r-r'|}{\lambda}\right) - N_0\left(\frac{|r-r'|}{\lambda}\right)\right)\right\}, \quad (7)$$

where $\lambda$ is the screening length, $H_0(x)$ is the Struve function and $N_0(x)$ is the Bessel function of second kind.

Once $H, \Sigma_S, \Sigma_D$ and $\Sigma_{int}$ are known, the Green's function ($G$) can be calculated from Eq. (1) and all the other quantities of interest (including conductance of the QPC) can be found following the procedure outlined in ref.[2]. For an impurity located at location $(x_1, y_1)$, we model its potential energy in the 2 DEG as follows,

$$U_{impurity}(x,y) = \frac{q^2}{4\pi\varepsilon_0\varepsilon_r\sqrt{(x-x_1)^2 + (y-y_1)^2 + \Delta^2}}, \quad (8)$$

where $\Delta = \frac{q}{4\pi\varepsilon_0\varepsilon_r U_0}$, and $U_0$ is the maximum strength of the impurity potential.

For an attractive impurity, we use the above expression with the negative sign. We use $\varepsilon_r = 12.9$, the relative dielectric constant of GaAs. In all simulations, the source potential $V_s = 0V$, and the potential at the drain contact $V_d = 0.1mV$. An asymmetry in the QPC potential confinement is introduced by taking $V_{sg1} = 0.2\text{ V} + V_{sweep}$ and $V_{sg2} = -0.2\text{ V} + V_{sweep}$ and the conductance of the constriction was studied as a function of the sweeping (or common mode) potential, $V_{sweep}$. The conductance of the QPC was then calculated using the NEGF with a non-uniform grid configuration containing more grid points at the interface of the QPC with vacuum. All calculations were performed at a temperature $T = 4.2\ K$. The screening length $\lambda$ in Eq.(1) was set equal to 5 nm and assumed independent of the gate potentials.

Hereafter, we also calculate the spin conductance polarization $\alpha = [G_\uparrow - G_\downarrow]/[G_\uparrow + G_\downarrow]$, where $G_\uparrow$ and $G_\downarrow$ are the conductance due to the majority and minority spin bands, respectively. We study how the maximum of alpha is affected by the strength, polarity, and location of the impurity potential.

## III. RESULTS

Figure 2 is a plot of the conductance as a function of $V_{sweep}$ for a QPC containing an impurity at location 3 in Fig.1. Also shown for comparison is the conductance plot with no impurity in the channel. The conductance anomalies are highly sensitive to the strength and type of impurity (attractive and repulsive). All curves show a conductance anomaly around 0.5 $G_0$ followed by a negative differential region (NDR) and a second anomaly somewhere between 0.5-1 $G_0$. The peak-to-valley ratio of the NDR after the 0.5 $G_0$ increases with the strength of the impurity potential for attractive impurity potential. The opposite trend is observed for a repulsive impurity. Figure 2 shows that there is a substantial shift of the conductance along the common signal $V_{sweep}$ for an impurity which is either attractive or repulsive. For comparison, we also show the conductance results for the case of no impurity in the channel.

In QPC experiments, the charge state of an impurity is often affected during sample handling, such as temperature cycling when the sample is brought back to room temperature between low temperature measurements. This typically leads to markedly different conductance traces for identical biasing conditions. Recently, we have observed this phenomenon while studying the conductance of InAs based QPCs in the presence of LSOC while varying the asymmetric bias applied between the two side gates. Thermal cycling is expected to change the charge state either remote impurities used in the modulation doping to form the 2DEG, or in the charge state of dangling bonds formed on the side walls of the QPC during etching, or even due to a single impurity located in the path of current flow.

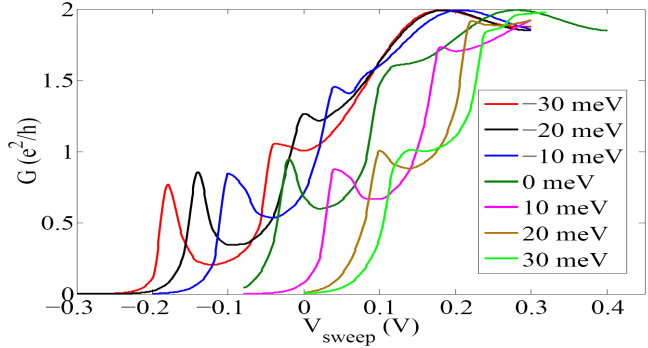

Fig. 2: Conductance as a function of $V_{sweep}$ for a QPC containing an impurity (either attractive or repulsive) at location 3 in Fig.1. Also shown for comparison are the results with no impurity (0 meV) in the channel.

The sensitivity of the conductance anomalies to the impurity location (points 1,2,3 in Fig.1) in the central portion of the QPC is illustrated in Fig.3.

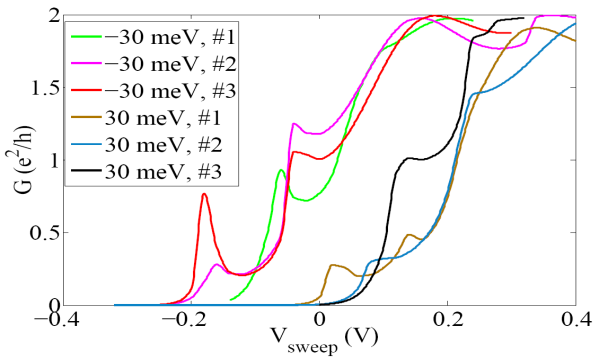

**Fig. 3:** Conductance as a function of $V_{sweep}$ for a QPC containing an impurity (either attractive or repulsive) at locations 1, 2, and 3 in Fig.1.

These results illustrate the strong sensitivity of the conductance anomalies on the impurity location.

The maximum value of α was found to be located near the first maximum in the conductance plots in Figures 2 and 3. In Fig.1, $α_{max}$ changes by 6% (from 0.989 to 0.927) when Δ varies from -30 to 30 meV. The change in $α_{max}$ is only 4% (from 0.976 to 0.948) when Δ changes from -20 to 20 meV, and less than 1% (from 0.976 to 0.97) when Δ changes from -10 to 10 meV. This dependence of $α_{max}$ on the type and strength of an impurity, even though by just a few percent, could lead to a substantial reduction in the ON/OFF conductance ratio of a spin valve built of two QPCs in series. Interestingly, the value of $α_{max}$ is larger (smaller) in the presence of an attractive (repulsive) impurity compared to the case of no impurity in the channel, for which $α_{max}$ = 0.973. To reach a larger value of $α_{max}$, a tunable repulsive impurity potential could be generated through the use of a negatively biased STM tip on top of the narrow portion of the QPC.

In Fig.4, we plot the conductance versus $V_{sweep}$ for a QPC containing two repulsive impurities located at $(x_1,y_1)$ = (16nm,28nm) and $(x_2,y_2)$ = (32nm,28nm), i.e., slightly off center. In this case, a reverse polarity was used, i.e., $V_{sg1}$ = -0.2 V + $V_{sweep}$ and $V_{sg2}$ = 0.2 V + $V_{sweep}$, In this case, the overall potential energy in the narrow portion of the QPC is larger closer to gate 1 and near the conductance anomaly, the spin down electrons are the majority carriers. We found $α_{max}$ = -0.956, a still rather large value despite the two impurities in the channel.

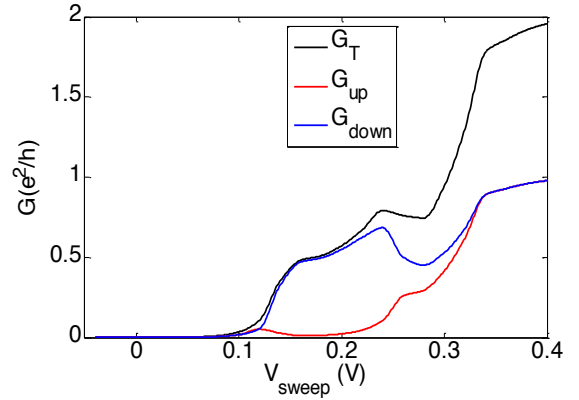

Fig. 4: Conductance as a function of $V_{sweep}$ for a QPC containing two repulsive impurities located at $(x_1,y_1)$ = (16,nm,28nm) and $(x_2,y_2)$ = (32nm,28nm). We used $V_{sg1}$ = -0.2 V + $V_{sweep}$ and $V_{sg2}$ = 0.2 V + $V_{sweep}$.

Finally, Fig.5 illustrates that, even for the case of symmetric bias on the two gates ($V_{sg1}$ = $V_{sg2}$ = 0.0 V + $V_{sweep}$) a conductance anomaly is found with a QPC with the same two off-center repulsive impurities as in the previous figure. In this case, the potential energy is also closer to gate 1 in the narrow portion of the QPC and the spin-down electrons are the majority carriers in the channel near the conductance anomaly. This last simulation shows that even a slight asymmetry due to unwanted impurities can lead to spin polarization in an otherwise perfectly symmetric channel.

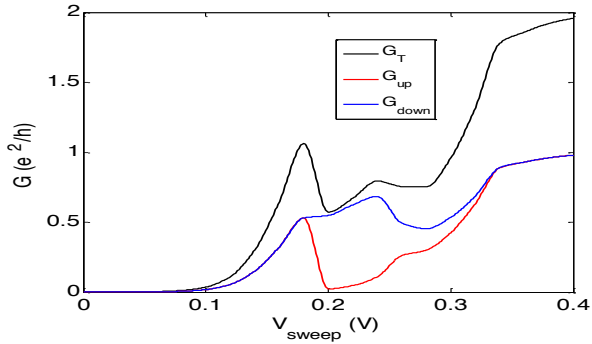

Fig. 5: Conductance as a function of $V_{sweep}$ for a QPC containing two repulsive impurities impurities located at $(x_1,y_1)$ = (16,nm,28nm) and $(x_2,y_2)$ = (32nm,28nm) for the case of a symmetrical bias condition, i.e., $V_{sg1} = V_{sg2}$ = 0.0 V + $V_{sweep}$.